\documentclass{emulateapj}
\usepackage{multirow}
\usepackage{subfigure}
\def\msun{$M_{\odot}$}

\def\xmm{{\it XMM-Newton}}

\shortauthors{Lin et al.}
\begin{document}

\title{Discovery of an Ultrasoft X-ray Transient Source in the 2XMM Catalog: a Tidal Disruption Event Candidate}

\author{Dacheng Lin\altaffilmark{1}, Eleazar R. Carrasco\altaffilmark{2}, Dirk Grupe\altaffilmark{3}, Natalie A. Webb\altaffilmark{1}, Didier Barret\altaffilmark{1},   and  Sean A. Farrell\altaffilmark{4}}
\altaffiltext{1}{Institut de Recherche en Astrophysique et Plan\'{e}tologie, 9, av du Colonel Roche, BP 44346, 31028 Toulouse Cedex 4, France, email: Dacheng.Lin@cesr.fr}
\altaffiltext{2}{Gemini Observatory/AURA, Southern Operations Center, Casilla 603, La Serena, Chile}
\altaffiltext{3}{Department of Astronomy and Astrophysics, Pennsylvania State University, 525 Davey Lab, University Park, PA 16802, USA}
\altaffiltext{4}{Sydney Institute for Astronomy (SIfA), School of Physics, The University of Sydney, NSW 2006, Australia}

\begin{abstract}
We have discovered an ultrasoft X-ray transient source,
\object{2XMMi~J184725.1-631724}, which was detected serendipitously in
two \xmm\ observations in the direction of the center of the galaxy IC
4765-f01-1504 at a redshift of 0.0353. These two observations were
separated by 211 days, with the 0.2--10 keV absorbed flux increasing
by a factor of about 9. Their spectra are best described by a model
dominated by a thermal disk or a single-temperature blackbody
component (contributing $\gtrsim$80\% of the flux) plus a weak
power-law component. The thermal emission has a temperature of a few
tens of eV, and the weak power-law component has a photon index of
$\sim$3.5. Similar to the black hole X-ray binaries in the thermal
state, our source exhibits an accretion disk whose luminosity appears
to follow the $L\propto T^4$ relation. This would indicate that the
black hole mass is about 10$^5$--10$^6$ \msun\ using the best-fitting
inner disk radius. Both \xmm\ observations show variability of about
21\% on timescales of hours, which can be explained as due to fast
variations in the mass accretion rate. The source was not detected by
{\it ROSAT} in an observation in 1992, indicating a variability factor
of $\gtrsim$64 over longer timescales. The source was not detected
again in X-rays in a {\it Swift} observation in 2011 February,
implying a flux decrease by a factor of $\gtrsim$12 since the last
\xmm\ observation. The transient nature, in addition to the extreme
softness of the X-ray spectra and the inactivity of the galaxy implied
by the lack of strong optical emission lines, makes it a candidate
tidal disruption event. If this is the case, the first \xmm\
observation would have been in the rising phase, and the second one in
the decay phase.

\end{abstract}


\section{INTRODUCTION}
\label{sec:intro}

It is generally believed that supermassive black holes (SMBHs) reside
in many galaxies, but only a fraction of them exhibit active galactic
nuclei (AGN), while others are dormant \citep{kori1995}.  Tidal
disruption events can provide a unique way to find and study the
dormant SMBHs \citep{re1988}. Such an event occurs when a star
approaches a SMBH and is tidally disrupted and subsequently accreted
\citep{lioz1979,re1988}. The mass of the SMBH should be
$\lesssim$$10^{8}$ \msun\ for such events to occur outside the event
horizon for solar-type stars. Tidal disruption events are expected to
be transient and rare, with the average occurrence rate of
$\sim10^{-4}$ yr$^{-1}$ per galaxy \citep{re1990}. They are predicted
to have a fast rise, with a timescale of half a year, and the decay
can last on the order of months to years, with the luminosity decaying
as $L\propto t^{-5/3}$ \citep{lioz1979,re1988,re1990}. The peak of the
flare is expected to reach the Eddington luminosity and be dominated
by thermal UV or X-ray emission.

A few tidal disruption event candidates were found from the {\it
ROSAT} All-Sky Survey, such as \object{RX J1624.9+7554}, \object{RX
J1242.6-1119} and \object{NGC 5905}
\citep{grthle1999,kogr1999,koba1999}. Their host galaxies were
confirmed to be inactive or only weakly active (\object{NGC 5905})
using {\it Hubble Space Telescope} spectroscopy
\citep{gehako2003}. They had peak soft X-ray luminosities up to
$\sim$10$^{44}$ erg s$^{-1}$ and showed ultrasoft X-ray spectra with
blackbody temperatures of $\sim$0.04--0.1 keV
\citep{ko2002,ko2008}. \object{NGC 5905} is the best observed
candidate and is the first one found to follow approximately the
$L\propto t^{-5/3}$ evolution. Some candidates were detected recently
from the \xmm\ Slew Survey \citep{essafr2007,essako2008} in the X-rays
and from the {\it Galaxy Evolution Explorer} Deep Imaging Survey in
the UV \citep{gemami2006,gebama2008,gehece2009}. Very recently, the
transient source \object{Swift J164449.3+573451} was suspected to be
due to a tidal disruption event \citep{blgime2011, bukegh2011}. In
contrast with the candidates above and the theoretical prediction,
this source is hard in X-rays, with a photon index around 1.8
\citep{bukegh2011}. Its tidal disruption event explanation still needs
to be confirmed by future long-term monitoring.

We are carrying out a project of classifying a sample of sources in
the Second \xmm\ Serendipitous Source (2XMM) Catalog
\citep{wascfy2009}. Here we report on the discovery of an ultrasoft
X-ray transient source, \object{2XMMi~J184725.1-631724}, whose
position is RA=18:47:25.16, Dec=-63:17:24.96 (J2000) from the 2XMM
catalog, with a 1-$\sigma$ error of 0\farcs26. It is in the direction
of the center of the galaxy IC 4765-f01-1504 \citep{camein2006}. This
source has negligible emission above 2 keV. We describe the
multi-wavelength observations of the source and the data reduction in
Section~\ref{sec:reduction}. In Section~\ref{sec:results}, we first
give the multi-wavelength detections of the source, followed by
presentations of its detailed X-ray spectral and timing properties. We
discuss its possible nature in Section~\ref{sec:discussion} and draw
our conclusions in Section~\ref{sec:conclusion}.

\section{DATA ANALYSIS}
\label{sec:reduction}

\subsection{\xmm\ Observations}

\tabletypesize{\scriptsize}
\setlength{\tabcolsep}{0.03in}
\begin{deluxetable*}{lcccccccc}
%\addtolength{\tabcolsep}{-5pt}
\tablecaption{\xmm\ Observation Log\label{tbl:obslog}}
\tablewidth{0pt}
\tablehead{ \colhead{Observation ID} &\colhead{Date}&\colhead{off-axis angles (arcmin)}   &\colhead{Duration} &\colhead{Exposure(ks)} &\colhead{Filter}   \\
               &               &\colhead{pn/MOS1/MOS2}&\colhead{(ks)}&\colhead{pn/MOS1/MOS2} &
}
\startdata
0405550401(XMM1) & 2006-09-06.98 &3.9/3.0/3.7 &28.0&19.5/27.6/27.6 &medium \\
0405380501(XMM2) & 2007-04-16.31 &9.0/8.5/9.4 &34.7&20.5/32.2/31.9 &thin1
\enddata 
\end{deluxetable*}

\object{2XMMi~J184725.1-631724} was observed twice by \xmm\
(Table~\ref{tbl:obslog}), on 2006 September 7 and 2007 April 16. These
two observations of this source will be referred to hereafter as XMM1
and XMM2, respectively. The source was detected in all the three
European Photon Imaging Cameras in the imaging mode, i.e., pn, MOS1,
and MOS2 \citep{jalual2001,stbrde2001,tuabar2001}, in both
observations. The source was also detected by the Optical Monitor
\citep[OM;][]{mabrmu2001} in XMM1, but it was not in the FOV of the OM
in XMM2. In XMM1, the two UV filters UVW1 and UVM2 were used, and we
obtained the source detection information directly from the pipeline
products.

We used SAS 10.0.0 and the calibration files of 2010 November for
reprocessing the X-ray event files and follow-up analysis. The data in
strong background flare intervals, mostly at the end of the XMM2
observation in the pn camera, are excluded following the SAS thread
for the filtering against high backgrounds. The final exposures used
are given in Table~\ref{tbl:obslog}. We extracted the source spectra
of the pn, MOS1, and MOS2 cameras from a circular region centered on
the source using 15$''$ and 35$''$ radii for XMM1 and XMM2,
respectively. A smaller radius was used for XMM1 because the source
was fainter and near the CCD gap. The background spectrum was
extracted from a large circular region with a radius of 100$''$ near
the source in each camera. The event selection criteria followed the
default values in the pipeline (see Table~5 in \citet{wascfy2009}). We
rebinned the spectra to have at least 20 counts in each bin so as to
adopt the $\chi^2$ statistic for the spectral fits.

We also extracted light curves from the pn camera, which has a larger
effective area and a higher timing resolution than the MOS cameras,
using the same apertures as those for spectral extraction. We first
extracted background-subtracted light curves with a bin size of 250~s,
using the SAS task {\it epiclccorr} to apply relative corrections. To
create the power density spectra (PDS), we also extracted light curves
from the source region using the frame time as the bin size, which is
199.1 ms for XMM1 (using the extended-full-frame mode) and 73.4 ms for
XMM2 (using the full-frame mode). Considering that the source is very
soft and the background dominates above 2 keV, all light curves were
extracted in the energy range 0.2--2.0 keV. We calculated the PDS
using a similar procedure as, e.g., \citet{gorore2006}. The XMM1 199.1
ms and XMM2 73.4 ms pn light curves were split into segments each with
32768 and 65536 data bins, respectively, resulting in four segments
for XMM1 and five for XMM2. The PDS was calculated for each segment,
and all PDS for each light curve were merged and averaged by binning
in frequency using a logarithmic factor of 1.1, under the condition that
each bin contains at least 20 individual PDS measurements. The errors
were calculated from the sample standard deviation of PDS measurements
in each bin.

\subsection{{\it ROSAT} and {\it Swift} Observations}
Our source was not detected in the {\it ROSAT} All-Sky Survey in 1990,
which had a detection limit of 0.1--2.4 keV flux 5$\times$10$^{-13}$
erg s$^{-1}$ cm$^{-2}$ \citep{voasbo1999}. Our source was in the FOV
of one {\it ROSAT} PSPC pointed observation (the sequence number
800256, 1992 October, $\sim$11 ks), at an off-axis angle of
$\sim$2.6$'$. It was not detected either and was (thus) not listed in
the WGA catalog of the {\it ROSAT} point sources
\citep{whgian1994}. We calculated the confidence interval of the
source detection using Bayesian statistics as described in
\citet{krbuno1991}. Circular source and background regions with radii
of 40$\arcsec$ and 2$\arcmin$ respectively were used. The
corresponding (ancillary plus photon redistribution) response matrix
was generated and used to convert the count rates to the fluxes.

At our request, the {\it Swift} Gamma Ray Burst Explorer mission
\citep{gechgi2004} observed the field of
\object{2XMMi~J184725.1-631724} on 2011 February 23 for a total of 5
ks (observation ID 00031930001). The X-ray telescope
\citep[XRT;][]{buhino2005} was operated in Photon Counting mode
\citep{hibuno2004}. X-ray data were reduced with the task {\it
xrtpipeline} version 0.12.1. We found an enhanced count rate at the
position of our source, but it is very weak. We also calculated the
confidence interval of the detection. Radii of 23\farcs5 and
235$\arcsec$ were used for the circular source and background regions,
respectively. The corresponding response matrix was generated using
the calibration files of 2011 February. The UV-Optical Telescope
\citep[UVOT;][]{rokema2005} was operated using the UVW1 filter for 5
ks. The magnitude and flux were measured with the task {\it
uvotsource} version 3 based on the most recent UVOT calibration as
described in \citet{pobrpa2008} and \citet{becuho2010}. Circular
source and background regions with radii of 5$\arcsec$ and
20$\arcsec$, respectively, were used.

\subsection{Optical Observations}

Our source is in the direction of the center of the galaxy IC
4765-f01-1504 \citep{camein2006}. This galaxy is located in the
background of the rich group of galaxies IC 4765 (also known as Abell
S0805, $z$=0.01497). It was imaged with the 1.3 m Warsaw telescope at
Las Campanas Observatory in Chile through the standard Johnson V and
Cousins I filters in 1998. We used the V- and I-filter images from
\citet{camein2006} to derive the main photometric parameters of the
galaxy with a S\'{e}rsic model in GALFIT \citep{pehoim2010}. The
images have a FWHM of the PSF of about 1\farcs2.

\citet{camein2006} also obtained an optical spectrum of the galaxy on
1999 June 19 with the Wide Field CCD camera mounted on the 2.5 m Du
Pont Telescope at the Las Campanas Observatory in Chile, but it has
poor quality. We obtained a new longslit spectrum of this galaxy with
the Gemini Multi-Object Spectrograph \citep[GMOS,][]{hojoal2004} at
the Gemini South Telescope in the queue mode. The observation was made
on the night of 2011 March 19 (UT) during bright time (illumination
fraction 0.99), under photometric conditions and $\sim$1\arcsec\
seeing. The 400 lines/mm ruling density grating (R400) centered at
5500\AA\ was used, to minimize the effect of moon illumination. The
slit width was set to 1$\arcsec$. A total of four exposures of 900 s
each were obtained. Small offsets in the spectral direction (50\AA)
towards the blue and the red were applied between exposures to allow
for the gaps between CCDs and to avoid any loss of important lines
present in the spectra.  Spectroscopic flats and comparison lamp
(CuAr) spectra were taken after each science exposure. In addition,
the spectrophotometric standard star LTT 7379 was observed at the end
of the night to flux calibrate the science spectrum.

The observations were processed with the Gemini IRAF package version
1.9 in IRAF. All science exposures, comparison lamps and spectroscopic
flats were bias subtracted and trimmed. Spectroscopic flats were
processed by removing the calibration unit plus GMOS spectral response
and the calibration unit uneven illumination, normalizing and leaving
only the pixel-to-pixel variations and the fringing. The resulting
two-dimensional spectra were then wavelength calibrated, corrected by
S-shape distortions, sky-subtracted, extracted to a one-dimensional
format using a fixed aperture of 7\farcs8 in diameter, and then
average combined. The final spectrum has a resolution of $\sim$8.8
\AA\ (FWHM) and a dispersion of $\sim$1.36 \AA\ pixel$^{-1}$, covering a
wavelength interval of $\sim$4000--7600~\AA. The signal-to-noise ratio
is about 40 at 5500 \AA.

We measured the redshift of the galaxy with two methods. In the first
method, we cross-correlated the spectrum with a high signal-to-noise
template using the {\it fxcor} routine in the IRAF RV package. The
error was estimated using the R statistic of \citet{toda1979}:
$\sigma_{v}=(3/8)(w/(1+R))$, where $w$ is the FWHM of the correlation
peak and $R$ is the ratio of the correlation peak height to the
amplitude of the antisymmetric noise. In the second method, we
identified the most prominent absorption lines (as no clear emission
lines were detected) in the spectrum and derived the redshift by
employing a line-by-line Gaussian fit using the {\it rvidline} routine
in the IRAF RV package.

\section{RESULTS}
\label{sec:results}

\subsection{The Source and the Multi-wavelength Observations}
\label{sec:srcandobs}

\tabletypesize{\scriptsize}
\setlength{\tabcolsep}{0.03in}
\begin{deluxetable*}{lcccccccccccc}
%\addtolength{\tabcolsep}{-5pt}
\tablecaption{The counterpart candidates in UV, optical, and IR\label{tbl:counterpart}}
\tablewidth{0pt}
\tablehead{\multicolumn{3}{c}{UV (XMM1 OM)}                                  &&\multicolumn{4}{c}{Optical (USNO B1.0)}&&\multicolumn{4}{c}{IR (2MASS PSC)}\\
\cline{1-3} \cline{5-8} \cline{10-13}
\colhead{$r$} & \colhead{UVW1} & \colhead{UVM2}                                             &&\colhead{$r$} & \colhead{B2}&\colhead{R2}&\colhead{I} &&\colhead{$r$} &\colhead{J}&\colhead{H}&\colhead{K}\\
\colhead{(arcsec)} &\multicolumn{2}{c}{(AB mag/flux($10^{-16}$ erg s$^{-1}$ cm$^{-2}$ \AA$^{-1}$))} &&\colhead{(arcsec)} &\multicolumn{3}{c}{(mag)}                &&\colhead{(arcsec)} &\multicolumn{3}{c}{(mag)}
}
\startdata
  0.52 & $19.65$$\pm$0.10/1.78$\pm$0.17 & $19.83$$\pm$$0.19$/2.38$\pm$0.41 &&0.16& 16.1 &15.33 &15.40  && 0.31&   15.29$\pm$0.07 & 14.59$\pm$0.09 &14.30$\pm$0.08
\enddata 
\tablecomments{The $r$ columns are the offsets of the counterparts from \object{2XMMi~J184725.1-631724}. The magnitudes/fluxes are not corrected for the Galactic reddening.}
\end{deluxetable*}

\begin{figure} 
\centering
\includegraphics{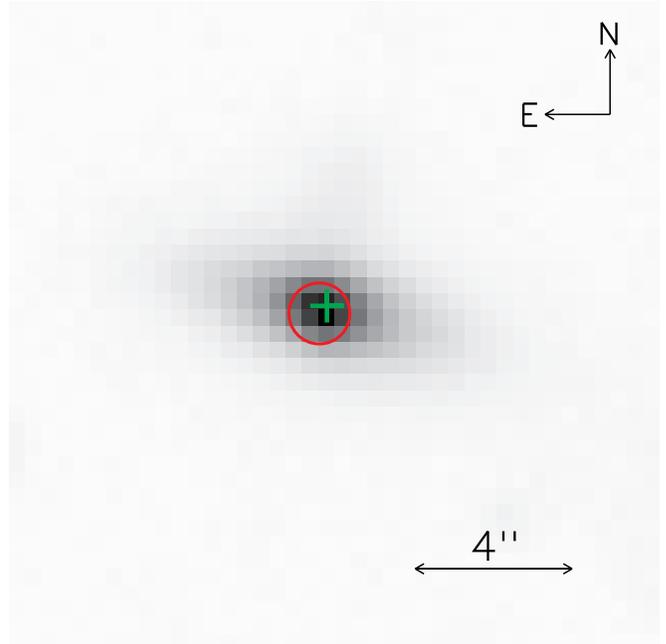}
\caption{The optical image of the galaxy in the V band. The pixel size
  is 0\farcs414. The green plus marks the central position of the
  galaxy obtained from the fits to its V- and I-band profiles using a
  S\'{e}rsic model, and it is at RA=18:47:25.14,
  Dec=$-$63:17:24.77 (J2000). The red circle is centered at the X-ray
  position, with the radius corresponding to the 3-$\sigma$
  error. \label{fig:optimg}}
\end{figure}
\begin{figure*}
\centering
\includegraphics[width=0.99\textwidth]{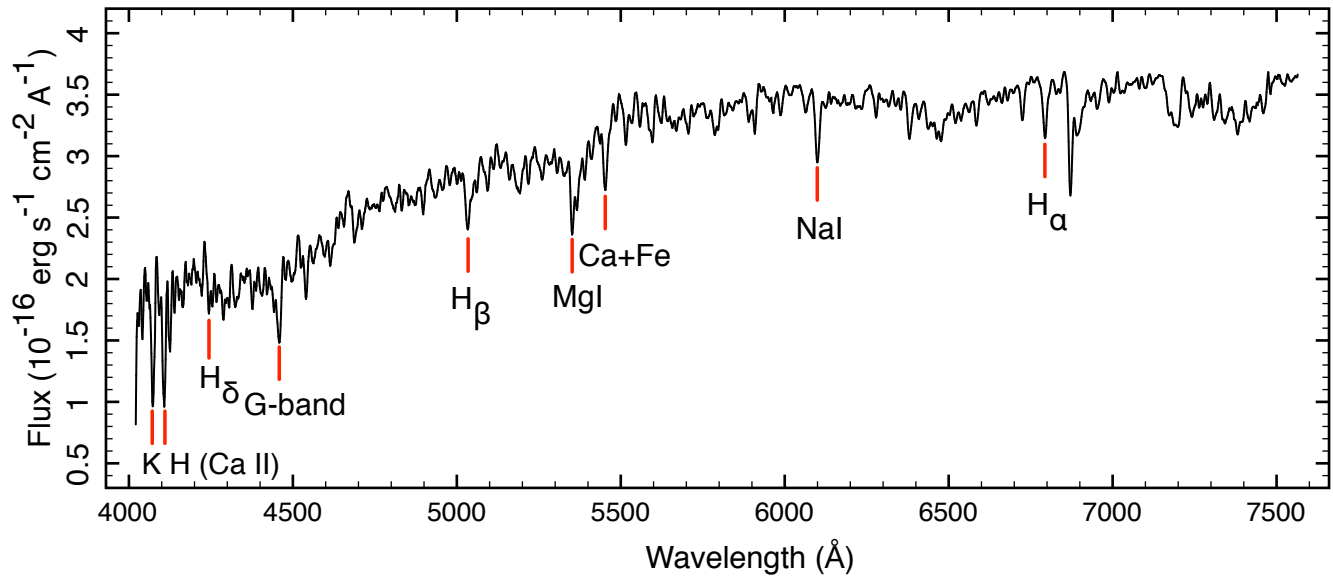}
\caption{The smoothed spectrum of the galaxy IC 4765-f01-1504 from the Gemini South Telescope, with the most important absorption lines identified. The drop at 6870 \AA\ is due to the atmosphere OH absorption. \label{fig:optspec}}
\end{figure*}

\object{2XMMi~J184725.1-631724} was detected in X-rays in both XMM1
and XMM2. We see no clear X-ray emission of our source from the {\it
ROSAT} observation in 1992 October and the {\it Swift} observation in
2011 February, indicating variability factors of $>$64 and $>$12,
respectively, compared with XMM2 (the 0.2--10 keV absorbed flux; see
Section~\ref{sec:rossw}). Here we concentrate on observations in other
wavelengths.

The source position from the 2XMM catalog has been astrometrically
corrected by matching with the optical catalog USNO B1.0
\citep{moleca2003}. For XMM1 and XMM2, a very small correction was
invoked (a fraction of an arcsec). We compared the corrected positions
of ten of the brightest X-ray sources in each observation with the
positions of optical counterparts from the USNO B1.0 catalog and found
that most of the offsets between the matches are less than 0.5$''$,
indicating successful astrometric corrections.

Table~\ref{tbl:counterpart} gives the detection of a UV source in UVW1
and UVW2 from the XMM1 OM at a position within the 2-$\sigma$
positional error from \object{2XMMi~J184725.1-631724} and is probably
its UV counterpart. There is also a UV source detected near our source
in UVW1 from the {\it Swift} UVOT. After applying the astrometric
correction using the USNO-B1.0 catalog with the \xmm\ SAS task {\it
eposcorr}, we obtain its position of RA=18:47:25.14 and
Dec=-63:17:25.04 (J2000), within the 1-$\sigma$ error from the X-ray
position. Its magnitude and flux are 18.67$\pm$0.04 and
(1.34$\pm$0.05)$\times$10$^{-16}$ erg~s$^{-1}$~cm$^{-2}$~\AA$^{-1}$,
respectively. We note that the UV filter set in the UVOT is different
from that of the \xmm\ OM, and the above values should not be directly
compared with the OM measurements in XMM1 above. \citet{grkoga2008}
measured an offset between the magnitudes from the two instruments of
$W1_{\rm OM} - W1_{\rm UVOT} = 0.78$ by comparing several field stars
in the images of the AGN Mkn 335. With this taken into account, there
seems to be little variability in the UV between the two epochs.

The optical and IR counterpart candidates of the source from the USNO
B1.0 and 2MASS Point Source Catalogs are given in
Table~\ref{tbl:counterpart}. The optical counterpart, the galaxy IC
4765-f01-1504, is shown in Figure~\ref{fig:optimg}
\citep{camein2006}. Our fits of the V- and I-band images using a
S\'{e}rsic model give integrated magnitudes of
16.99$\pm$0.01 and 15.62$\pm$0.02, effective radii of
2\farcs52$\pm$0\farcs04 and 2\farcs94$\pm$0\farcs09, S\'{e}rsic
indices of 3.54$\pm$0.06 and 4.14$\pm$0.13, and apparent axis ratios
of 0.238$\pm$0.003 and 0.271$\pm$0.003, respectively. This galaxy may
be an elliptical galaxy, which typically has a S\'{e}rsic index of
4. The axis ratios above would imply a high inclination of this galaxy
if its intrinsic ellipticity is low.

Figure~\ref{fig:optspec} shows the spectrum of IC 4765-f01-1504 from
the Gemini South Telescope. No clear emission lines were detected,
supporting the identification as an elliptical galaxy. A redshift of
$z$=0.0353$\pm$0.0001 was obtained from both the cross-correlation
method, with $R=12.80$, and the absorption line fit method, indicating
a perfect agreement between them. This redshift disagrees with the
value of 0.0869 obtained by \citet{camein2006}, which used the
cross-correlation method (as there were no significant emission lines
detected). Considering that our new spectrum has much better quality,
we adopt this new redshift. We measured a 3-$\sigma$ upper limit of
[OIII] 5007 \AA\ of 0.9$\times$$10^{-15}$ erg s$^{-1}$
cm$^{-2}$. Assuming a flat universe with the Hubble constant $H_0$=73
km s$^{-1}$ Mpc$^{-1}$ and the matter density $\Omega_{\rm M}$=0.27,
this redshift corresponds to a comoving radial distance of 143.9 Mpc
and a luminosity distance of 149.0 Mpc, which will be used in this
paper. The absolute V and K magnitudes of this galaxy are -19.2 and
-21.6, respectively, after the Galactic extinction correction
\citep{scfida1998}. Based on the BH mass vs. bulge luminosity
relations from \citet{gr2007}, \citet{lafari2007} and
\citet{mahu2003}, the above magnitudes imply the mass of the SMBH in
IC 4765-f01-1504 to be about 10$^7$ and 10$^6$ \msun\ if the
bulge/total luminosity ratio is 1 or 0.1, respectively. Because the
sample of galaxies in the above studies were generally brighter than
IC 4765-f01-1504 and these relations have large intrinsic scattering,
these mass estimates might have an uncertainty as large as one order
of magnitude.

\subsection{X-ray Spectral Modeling}
\label{sec:spmod}
\tabletypesize{\scriptsize}
\setlength{\tabcolsep}{0.03in}
\begin{deluxetable*}{lccccccccccc}
%\addtolength{\tabcolsep}{-5pt}
%\tabletypesize{\scriptsize}
\tablecaption{Spectral modeling results\label{tbl:mcd+pl}}
\tablewidth{0pt}
\tablehead{\colhead{Model} &\colhead{Obs} &\colhead{$N_{\rm H}$} & 
  \colhead{$kT_{\rm MCD/BB}$} &\colhead{$N_{\rm MCD/BB}$} &  
  \colhead{$\Gamma_{\rm PL/SIMPL}$} & \colhead{$N_{\rm PL}/f_{\rm SC}$} &
  \colhead{$\chi^2_\nu(\nu)$} & \colhead{$f_{\rm MCD/BB}$} & 
  \colhead{$F_{\rm abs}$} & \colhead{$F_{\rm unabs}$} & \colhead{$L$}\\
  & & \colhead{($10^{20}$~${\rm cm}^{-2}$)} &
  \colhead{(eV)} & \colhead{($10^{4}$)} &
  & \colhead{($10^{-5}/\%$)} & 
  &(\%) &\multicolumn{2}{c}{($10^{-12}$ erg s$^{-1}$ cm$^{-2}$)} & ($10^{43}$ erg s$^{-1})$
%\colhead{(1)} & \colhead{(2)} &\colhead{(3)} &\colhead{(4)} &\colhead{(5)} &\colhead{(6)} &\colhead{(7)} &\colhead{(8)} & \colhead{(9)}
}
\startdata
\multirow{2}{*}{MCD+PL} & XMM1 & \multirow{2}{*}{$8.6$$\pm$$ 0.5$} & $65.8$$\pm$$ 5.0$ & $1.56^{+1.21}_{-0.63}$ & $3.72$$\pm$$ 0.62$ & $1.79$$\pm$$ 0.32$ &0.92(115) & $84.5^{+7.3}_{-14.8}$ & $0.22$$\pm$$ 0.01$ & $1.76$$\pm$$ 0.19$ &$ 1.70$$\pm$$ 0.44$\\
&XMM2 & & $93.1$$\pm$$ 2.2$ & $1.45^{+0.34}_{-0.23}$ & $3.27$$\pm$$ 0.70$ & $3.75$$\pm$$ 1.39$ &1.20(284) & $96.4^{+1.9}_{-6.1}$ & $1.93$$\pm$$ 0.03$ & $10.18^{+1.00}_{-0.57}$ &$ 6.38$$\pm$$ 0.66$\\
\hline
\multirow{2}{*}{SIMPL(MCD)} &XMM1 & \multirow{2}{*}{$8.5$$\pm$$ 0.5$} & $65.1$$\pm$$ 5.5$ & $1.73^{+1.65}_{-0.73}$ & $3.71$$\pm$$ 0.63$ & $3.33^{+3.98}_{-1.69}$ &0.92(115) & $90.0^{+3.7}_{-7.1}$ & $0.22$$\pm$$ 0.01$ & $1.75^{+0.35}_{-0.24}$ &$ 1.81$$\pm$$ 0.57$\\
& XMM2 &  & $92.9$$\pm$$ 2.4$ & $1.49$$\pm$$ 0.32$ & $3.33$$\pm$$ 0.78$ & $0.83^{+1.32}_{-0.49}$ &1.20(284) & $97.7^{+0.9}_{-2.2}$ & $1.92$$\pm$$ 0.03$ & $10.21$$\pm$$ 0.89$ &$ 6.40$$\pm$$ 0.72$\\
\hline
\multirow{2}{*}{BB+PL} &XMM1 & \multirow{2}{*}{$7.4$$\pm$$ 0.7$} & $57.8$$\pm$$ 3.9$ & $1.63^{+1.13}_{-0.56}$ & $3.71$$\pm$$ 0.59$ & $1.76$$\pm$$ 0.28$ &0.92(115) & $78.2^{+9.5}_{-17.2}$ & $0.22$$\pm$$ 0.01$ & $1.23$$\pm$$ 0.11$ &$ 0.59$$\pm$$ 0.14$\\
& XMM2 & & $78.2$$\pm$$ 1.6$ & $2.37$$\pm$$ 0.41$ & $3.89$$\pm$$ 0.61$ & $5.54$$\pm$$ 1.46$ &1.18(284) & $86.0^{+7.6}_{-14.4}$ & $1.92$$\pm$$ 0.03$ & $7.59^{+1.15}_{-0.51}$ &$2.79$$\pm$$ 0.39$\\
\hline
\multirow{2}{*}{SIMPL(BB)} &XMM1 & \multirow{2}{*}{$7.2$$\pm$$ 0.5$} & $57.7$$\pm$$ 4.1$ & $1.81^{+1.3}_{-0.65}$ & $3.71$$\pm$$ 0.59$ & $5.48^{+4.81}_{-2.31}$ &0.92(115) & $87.8^{+3.9}_{-7.1}$ & $0.22$$\pm$$ 0.01$ & $1.18$$\pm$$ 0.18$ &$ 0.59$$\pm$$ 0.15$\\
& XMM2 &  & $77.5$$\pm$$ 2.2$ & $2.58^{+0.71}_{-0.45}$ & $4.25$$\pm$$ 0.83$ & $3.35^{+3.56}_{-1.63}$ &1.18(284) & $94.2^{+2.4}_{-4.6}$ & $1.92$$\pm$$ 0.03$ & $7.2$$\pm$$ 0.68$ &$ 2.70$$\pm$$ 0.31$
\enddata 
\tablecomments{The column $f_{\rm MCD/BB}$ refers to the unabsorbed flux fraction of the MCD/BB component in the 0.2--10 keV energy band (for the models SIMPL(MCD) and SIMPL(BB), it refers to the unscattered part). $F_{\rm abs}$ and $F_{\rm unabs}$ are the total absorbed and unabsorbed fluxes in the 0.2--10 keV energy band, respectively. The luminosity $L$ was calculated using the unabsorbed bolometric flux of each spectral component (the PL component was integrated down to 0.2 keV). All errors are at a 90\%-confidence level.}
\end{deluxetable*}

\begin{figure*}
\centering
\includegraphics[width=0.99\textwidth]{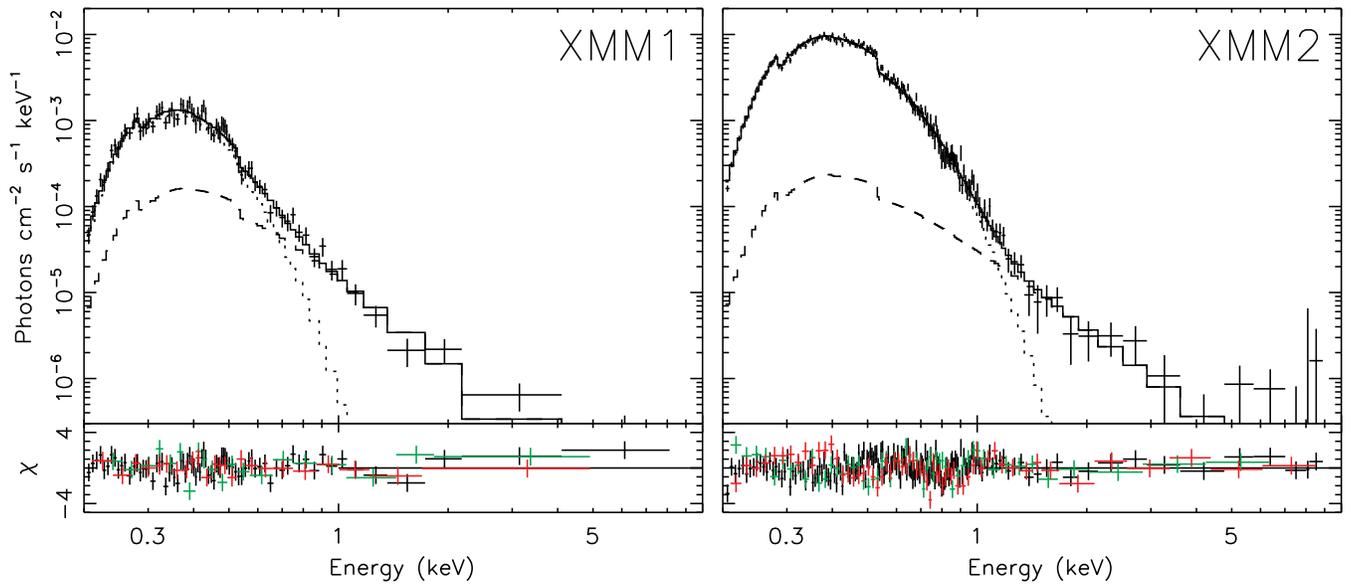}
\caption{The unfolded spectra and the fit residuals using the model MCD+PL. For clarity, only the pn spectra are shown for the unfolded spectra. The dotted, dashed, and solid lines are for the MCD and PL components and the total model, respectively. The residuals are shown for all three cameras (black/red/green for pn/MOS1/MOS2, respectively).\label{fig:spfits}}
\end{figure*}

\begin{figure}
\centering
\includegraphics{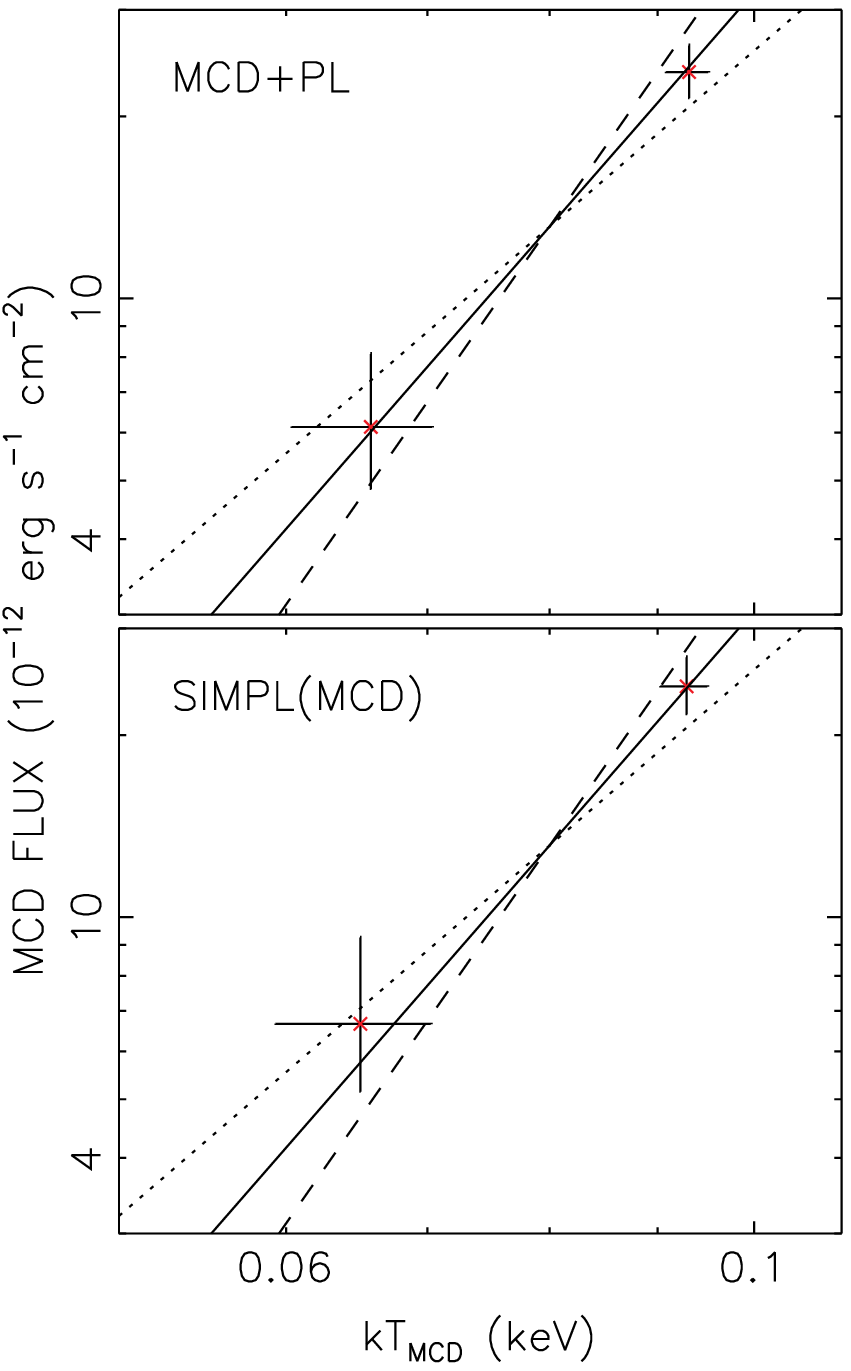}
\caption{The MCD flux versus $kT_{\rm MCD}$ using the models MCD+PL and SIMPL(MCD). The solid, dotted, and dashed lines plot the $L\propto T^4$, $L\propto T^3$, and  $L\propto T^5$ relations, as a guidance. \label{fig:spfits2}}
\end{figure}

We fitted the spectra of \object{2XMMi~J184725.1-631724} from both XMM1
and XMM2 using various spectral models. We jointly fitted the spectra
from all three cameras, i.e., pn, MOS1, and MOS2, and their relative
normalizations were left free. We only report the normalization results
corresponding to the pn camera. MOS1 and MOS2 differ by about 10$\%$
(the largest one $\sim$20\%), less than the error bars. We fitted the
spectra in the 0.2--10 keV energy band.

We were unsure the X-ray emission mechanism of our source. Thus we
first tested the common single-component models to see whether any of
them can describe our X-ray spectra well: a single temperature
blackbody (BB), a multi-color disk (MCD), a PL, a broken PL, a cut-off
PL, an APEC thermal plasma model, and a thermal bremsstrahlung
spectrum. They are models bbodyrad, diskbb, powerlaw, bknpower,
cutoffpl, APEC, and bremss in XSPEC, respectively. All models include
the absorption described by the WABS model in XSPEC; our results
change little with alternative absorption models such as PHABS or
TBABS in XSPEC. All these simple models fail to describe one or both
of the \xmm\ spectra, with residuals above 1 keV typically seen. For
indication, we report the PL index $\Gamma_{\rm PL}$ of the fits using
the PL model. We obtained $\Gamma_{\rm PL}=5.86$$\pm$0.27 for XMM1
($\chi^2_\nu$($\nu$)=1.13(117)) and 6.88$\pm$$0.09$ for XMM2
(($\chi^2_\nu$($\nu$)=2.25(286)). The lower $\chi^2$ value for XMM1 to
some degree is due to poorer data.

We next attempted to fit the spectra with the double-component models
MCD+PL and BB+PL, finding that they describe both the XMM1 and XMM2
spectra much better than the above common single-component models,
with the $\chi^2$ values decreased by more than 140 for the total
degrees of freedom of about 400 of both the XMM1 and XMM2 spectra. As
a way to model the hard component self-consistently, we also fitted
the spectra with SIMPL(MCD) and SIMPL(BB). SIMPL \citep[in
  XSPEC12;][]{stnamc2009} is an empirical convolution model of
Comptonization in which a fraction ($f_{\rm SC}$) of the input seed
photons are converted into a power law parametrized by an index
($\Gamma_{\rm SIMPL}$). We assume that all the scattered photons are
up-scattered in energy in this model.

The best-fitting values of the column density are consistent between
XMM1 and XMM2, with $N_{\rm H}$=$(7.6^{+1.5}_{-2.6})$ and
$(8.6$$\pm$$0.6)$$\times$$10^{20}$ cm$^{-2}$, respectively, using the
model MCD+PL. Thus, we chose to fit both spectra with a common value
of $N_{\rm H}$. The final results are given in
Table~\ref{tbl:mcd+pl}. The best-fitting values of $N_{\rm H}$ are
slightly higher than the Galactic value of 6.1$\times$$10^{20}$~${\rm
cm}^{-2}$ from the Leiden/Argentine/Bonn Survey of Galactic HI
\citep{kabuha2005}, probably indicating a small intrinsic absorption.

For the model MCD+PL, we show the unfolded spectra and residuals in
Figure~\ref{tbl:mcd+pl}. In this model, the spectra are dominated by
the MCD component at energies below 1 keV. The fraction of the MCD
component is about 84.5\% and 96.4\% for XMM1 and XMM2, respectively
(the 0.2--10 keV unabsorbed flux; Table~\ref{tbl:mcd+pl}). The 0.2--10
keV flux increases from XMM1 to XMM2 by a factor of 8.8 (absorbed) or
5.8 (unabsorbed). We also estimate the luminosity, using the
bolometric flux of each spectral component. The disk inclination is
uncertain, and we assume it to be 60$\degr$. The PL component diverges
at low energies, and we integrate its flux above 0.2 keV. We obtain
luminosities of 1.70 and 6.38$\times$10$^{43}$ erg s$^{-1}$ for XMM1
and XMM2, respectively (Table~\ref{tbl:mcd+pl}). For comparison, the
corresponding 0.2--10 keV luminosities are 0.47 and
2.70$\times$10$^{43}$ erg s$^{-1}$, respectively.

We plot the MCD bolometric flux versus its temperature at the inner
disk radius $kT_{\rm MCD}$ in Figure~\ref{fig:spfits2} (the upper
panel). We can see that the evolution of the MCD luminosity is
consistent with the $L\propto T^4$ track (the solid line), which
implies a constant inner disk radius with the change in luminosity. We
note that this is based on the only two observations available. The
disk temperature is relatively low, only $kT_{\rm MCD}=65.8$ and
$93.1$ eV for XMM1 and XMM2, respectively. The PL component is weak,
and its parameter values have relatively large uncertainties. Its
index is consistent between XMM1 and XMM2 and is relatively steep,
with $\Gamma_{\rm PL}$ about 3.5. Forcing XMM1 and XMM2 to have the
same value of $\Gamma_{\rm PL}$ in the fit, we see a change of the PL
normalization $N_{\rm PL}$ by a factor of 2.6 (4.5 $\sigma$).

The model SIMPL(MCD) gives results very similar to the model MCD+PL
(Table~\ref{tbl:mcd+pl} and Figure~\ref{fig:spfits2}), in terms of the
MCD temperature, the thermal fraction, etc. It infers that only about
$f_{\rm SC}$$=$$3\%$ and $1\%$ of the thermal disk emission is
Comptonized to the hard emission in XMM1 and XMM2, respectively. This
model, with a natural cutoff at low energies for the hard component,
infers luminosities similar to those of the model MCD+PL obtained by
integrating the PL flux down to 0.2 keV
(Table~\ref{tbl:mcd+pl}).

The spectra can also be fitted almost equally well using the models
BB+PL and SIMPL(BB) (Table~\ref{tbl:mcd+pl}). The BB component
dominates in both XMM1 and XMM2, contributing $\gtrsim$80\% of the
0.2--10 keV flux, similar to the MCD component in the models MCD+PL
and SIMPL(MCD). Its effective temperature $kT_{\rm BB}$ is also low,
about 58 and 78 eV for XMM1 and XMM2, respectively. The MCD and BB
models have very similar spectral shapes at high energies
\citep{mamami1986}, but their differences become large at low
energies. We estimate their differences in the UV. We have
measurements from two UV filters, i.e., UVW1 and UVM2, from XMM1. The
flux densities of the MCD component in the model MCD+PL from XMM1 are
($2.15$$\pm$$0.69$) and ($3.68$$\pm$$1.18$)$\times$10$^{-17}$ erg
s$^{-1}$ cm$^{-2}$ \AA$^{-1}$ at the effective wavelengths of UVW1
(2910 \AA) and UVM2 (2310 \AA), respectively. The corresponding values
for the BB component in the model BB+PL from XMM1 are
($0.40$$\pm$$0.16$) and ($1.00$$\pm$$0.40$)$\times$10$^{-19}$ erg
s$^{-1}$ cm$^{-2}$ \AA$^{-1}$, respectively. The corresponding flux
densities measured with UVW1 and UVW2 are ($3.00$$\pm$$0.47$) and
($5.08$$\pm$$1.45$)$\times$10$^{-16}$ erg s$^{-1}$ cm$^{-2}$
\AA$^{-1}$, respectively, after the Galactic dust extinction
correction using a reddening value of $E_{\rm (B-V)}=0.098$
\citep{scfida1998} and assuming a spectral shape of a MCD model at low
frequencies (i.e., a power law with a photon index of $2/3$). We see
that the UV flux from the OM detection is much higher than the BB flux
in the UV. It is closer to the MCD flux in the UV, but still there is
about an order of magnitude difference, which will be discussed in
Section~\ref{sec:discussion}. The flux of the PL component in the UV
is hard to assess as this model is too steep and diverges at low
energies (more than two orders of magnitude higher than that measured
by the OM) and must be cut off below some energy. The models
SIMPL(MCD) and SIMPL(BB) show no such problem, and their fluxes in the
UV, from the whole model or only from the thermal components, are very
close to the MCD and BB fluxes in the UV obtained above.

\subsection{Fast X-ray Variability}
\label{sec:pds}
\tabletypesize{\scriptsize}
\setlength{\tabcolsep}{0.03in}
\begin{deluxetable}{lcccccccc}
%\addtolength{\tabcolsep}{-5pt}
%\tabletypesize{\scriptsize}
\tablecaption{The fit results of the PDS using a PL plus a constant. \label{tbl:timing}}
\tablewidth{0pt}
\tablehead{\colhead{Obs} &
  \colhead{$\Gamma_{\rm PL}$} & \colhead{$N_{\rm PL}$ ($10^{-5}$)} &
  \colhead{$C_{\rm P}$} &
  \colhead{$\chi^2_\nu(\nu)$} & 
  \colhead{rms(\%)}
}
\startdata
XMM1 & $1.74^{+1.13}_{-0.57}$ & $3.49^{+27.70}_{-3.48}$ & 25.59$\pm$0.17 & 1.07(75) & 21.2$\pm$5.3 \\
XMM2 &1.65$\pm$0.19 &$8.17^{+20.42}_{-6.75}$ & 2.77$\pm$0.01 & 1.06(82) &21.1$\pm$1.9
\enddata 
\tablecomments{The column rms refers to 0.0001--0.01 Hz fractional rms after subtracting the Poisson level. $N_{\rm PL}$ is the PL normalization at 1 Hz. All errors are at a 90\%-confidence level, except for the rms, whose 1-$\sigma$ errors are given.}
\end{deluxetable}

\begin{figure*} 
\centering
\includegraphics{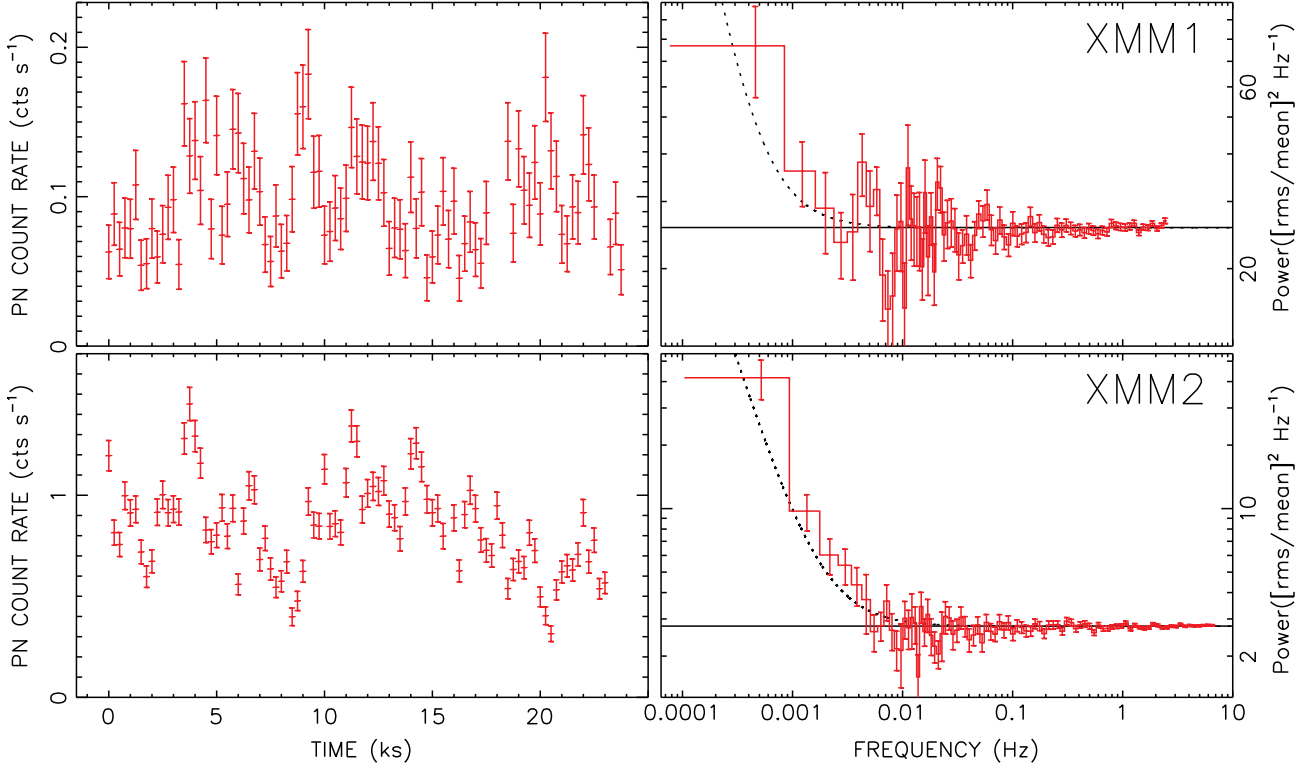}
\caption{Left panels: The pn 250~s 0.2--2.0~keV background-subtracted light curves. Right panels: The PDS of the pn 0.2--2.0 keV light curves binned at 199.1 ms for XMM1 and 73.4 ms for XMM2. The black solid constant line is the average PDS above 0.1 Hz, representing the Poisson level, and the dotted line is the best-fitting model of a PL plus a constant. \label{fig:lcfft}}
\end{figure*}

The left panels of Figure~\ref{fig:lcfft} show the pn 250 s
background-subtracted light curves. The background is at a level of
about 3$\%$ and 2$\%$ for XMM1 and XMM2, respectively. The variations
of the source count rate can be clearly seen for both
observations. For XMM2, which has higher count rates, we see that the
source varies by a factor of $\sim$4 within 5 ks.

The right panels of Figure~\ref{fig:lcfft} show the pn PDS of XMM1 and
XMM2. Both PDS are flat at frequencies above $0.1$ Hz, representing
the Poisson level. Their averages above $0.1$ Hz weighted by the
errors differ from the expected Poisson noise values by
$<$$0.3\%$. Below 0.01 Hz, both PDS show a clear deviation from the
Poisson level (the black solid line). We successfully fitted both PDS
with a power-law (PL) plus a constant $C_{\rm P}$, accounting for the
Poisson level, and the results are given in
Table~\ref{tbl:timing}. The PL index $\Gamma_{\rm PL}$ is about 1.7
for both observations. We evaluate the fractional rms within 0.0001 to
0.01 Hz after subtracting the Poisson level and obtain a value of
(21.2$\pm$5.3)\% and (21.1$\pm$1.9)\% for XMM1 and XMM2,
respectively. The above results indicate that the source shows similar
variability in both XMM1 and XMM2.

The large fast variability seen above should be due to the thermal
component from the models in Table~\ref{tbl:mcd+pl}, as the hard
component is very weak (e.g., $<$4\% in XMM2, in terms of the 0.2--2.0
keV pn count rates). To see the cause of the variability, we extracted
the high- and low-state spectra from the bright observation XMM2,
corresponding to intervals with the pn count rate higher or lower than
0.8 counts~s$^{-1}$, respectively (Figure~\ref{fig:lcfft}). We fitted
these two spectra simultaneously using the model SIMPL(MCD) with a
common value of $N_{\rm H}$. We find that the MCD temperature in the
low state is smaller than in the high state by 10 eV at a 4.4-$\sigma$
confidence level. Their MCD normalizations are consistent with being
the same within the error bars. The parameters of the hard component
have large uncertainties, making it hard to constrain any trend. Based
on the MCD model, the above results provide one explanation of our
large fast variability: it is caused by the fast variations in the
mass accretion rate, with the disk truncated at a constant radius.

\subsection{Comparison with ROSAT and Swift Observations}
\label{sec:rossw}
We obtain count rates of $<$1.9$\times$10$^{-3}$ and
1.3$^{+2.4}$$\times$10$^{-3}$ counts s$^{-1}$ for the {\it ROSAT} PSPC
observation in 1992 October and the {\it Swift} XRT observation in
2011 February, respectively. The 3-$\sigma$ upper bounds are given
above, and the lower bounds are zero. Using the response matrices
corresponding to the source extraction regions used and the fit
results of the model SIMPL(MCD) for XMM1, these count rates correspond
to the 0.2--10 keV absorbed fluxes of $<$0.3 and
0.6$^{+1.0}$$\times$$10^{-13}$ erg s$^{-1}$ cm$^{-2}$, respectively,
and the 0.2--10 keV unabsorbed fluxes of $<$2.4 and
4.5$^{+8.3}$$\times$$10^{-13}$ erg s$^{-1}$ cm$^{-2}$,
respectively. Thus, the source varied by a factor of $>$64 and $>$43
between the {\it ROSAT} pointed observation in 1992 and XMM2, using
the 0.2--10 keV absorbed and unabsorbed fluxes, respectively. The
corresponding variation factors between XMM2 and the {\it Swift}
observation in 2011 are $>$12 and $>$8, respectively.

\section{DISCUSSION}
\label{sec:discussion}

\subsection{The Nature and Implication of the Soft Component}

The remarkable features of \object{2XMMi~J184725.1-631724} are the
extreme softness of its X-ray spectra and the large
variability. Understanding the nature and implication of the soft
component which dominates the X-ray spectra will help to pin down the
nature of the source. As it is in the direction coincident with the
center of the galaxy IC 4765-f01-1504, we first assume that its X-ray
emission is associated with the SMBH in this galaxy. This can be due
to either a tidal disruption event or an AGN.

The fits with the models MCD+PL and SIMPL(MCD) above assume that there
was a thermal disk emission, whose luminosity fraction was inferred to
be very high, $\sim$90\%. The evolution of the MCD luminosity is
consistent with $L\propto T^4$, though only two observations are
available. These properties are very similar to the thermal state of
BH X-ray binaries \citep{remc2006}, in which the accretion disk is
believed to be truncated at the innermost stable circular orbit
(ISCO). Thus we assume that the disk is also truncated at the ISCO
during these two observations and roughly estimate the BH mass from
the MCD normalization $N_{\rm MCD}$. Using a distance of 143.9 Mpc and
assuming a disk inclination of 60$\degr$, we infer the BH mass to be
$\sim$3$\times$$10^5$ \msun, neglecting factors such as the spin and
the hardening effect. If we replace the MCD model with the more
realistic accretion disk model around a Kerr black hole kerrbb
\citep[in XSPEC;][]{lizina2005} and explore the parameter spaces of
the disk inclination 0--75$\degr$, the spin parameter $a^*$ 0--1, and
the hardening factor 1--1.7, we obtain a range of the BH mass of
(0.06--3.81)$\times$10$^6$ \msun. Assuming the BH mass to be
5$\times$$10^5$ \msun, the source would be at about 0.3 and 1.0
Eddington luminosity in XMM1 and XMM2, respectively, common values
seen in the thermal state of BH X-ray binaries \citep{dogiku2007}. We
note that the light crossing time of the inner accretion disk around a
BH with this mass is $\sim$50 s, about the timescale on which the
source begins to show strong variability (Figure~\ref{fig:lcfft}). The
above mass estimate is consistent with that using the BH mass
vs. bulge luminosity relations (Section~\ref{sec:srcandobs}),
considering the large uncertainties of both methods.

Some AGN can be very soft, showing strong soft excesses
\citep{pumaco1992}. The soft excess refers to the excess of emission
below $\sim$2 keV with respect to the extrapolation of the power-law
fit of the continuum above 2 keV and is commonly observed in type-I
AGN \citep{tupo1989}. The strongest soft X-ray excesses and
variability are found in Narrow Line Seyfert 1 galaxies \citep[NLS1s;
e.g.,][]{bobrfi1996,le1999a,le1999b,grkole2010}. The nature of the
soft excess is still unclear. The above thermal disk model is one of
the several competing models invoked \citep[e.g.,][]{wafi1993}. The
problem with this explanation is that the soft excesses from a sample
of AGN with a large range of mass and luminosity have characteristic
temperatures spanning a narrow range ($\sim$0.1--0.2 keV), which is
hard to explain \citep{gido2004,crfaga2006}. The narrow range of the
characteristic temperatures of the soft excesses finds a natural
explanation if they are due to atomic processes. There are two main
scenarios, i.e., absorption and reflection. \citet{gido2004} proposed
the soft excess as an artifact of strong, relativistically smeared,
partially ionized absorption. We test this model using the swind1
model from XSPEC. We use the model swind1(PL), with $\Gamma_{\rm PL}$
required to be $<$3.5 to make sure that it is not the steep PL
describing the soft excess. We obtain the values of $\chi^2_\nu(\nu)$
to be 1.32(114) and 1.31(283) for XMM1 and XMM2, respectively. Both
observations require strong velocities ($>$0.5 and 0.29$\pm$0.02 speed
of light for XMM1 and XMM2, respectively) to smear the
absorption/emission lines, which is hard to achieve from a radiatively
driven accretion disk wind \citep{scdo2007,scdopr2009}. The difference
between XMM1 and XMM2 is mainly due to different absorption column
densities and smearing velocities, requiring dramatic changes in the
absorber.

In the reflection model, a series of soft X-ray emission lines below 2
keV, if strongly relativistically blurred, can produce the smooth soft
excess feature. We follow \citet{crfaga2006} to use the model
kdblur(PL+reflionx), where kdblur is the relativistic convolution and
reflionx is a table model of the ionized reflection (see their
references therein). In the fits we force XMM1 and XMM2 to have common
values of $N_{\rm H}$, the inclination, and the Fe abundance. We
obtain $\chi^2_\nu(\nu)$=0.95(111) and 1.34(281) and reflection flux
fractions $\sim$0.94 and 1 (0.2--10 keV, unabsorbed) for XMM1 and
XMM2, respectively. The fits require a low inclination
(17.1$\pm$7.6$\degr$), a high Fe abundance ($8.37_{-2.65}^{+0.79}$
solar value), and a steep illuminating power-law spectrum (reaching
the upper index limit of 3.3 allowed in the model). The disk
emissivity index is different between these two observations
($5.37_{-0.27}^{+0.87}$ and 8.99$\pm$0.89 for XMM1 and XMM2
respectively), implying a very different disk structure. This model
infers a highly spinning BH, with the inner disk radius at
2.38$\pm$0.38 and $1.87_{-0.19}^{+0.53}$ gravitational radii for XMM1
and XMM2 respectively. Compared with the results from
\citet{crfaga2006}, the above values are extreme but not unique.

We see that both the absorption and reflection models invoke extreme
environments to explain our source. The former requires absorbers at
very high velocities and varying dramatically. The latter requires
very steep illuminating spectra and very different disk emissivities
between XMM1 and XMM2. In comparison, the thermal disk
emission explanation for the soft component in our source is more
reasonable. In this model, the difference between XMM1 and XMM2 is
simply due to the change in the accretion rate. We note that the
inferred inner disk temperatures are lower than the typical
characteristic temperatures of the soft excesses in AGN, which might
indicate that the soft component of our source has a different nature
from soft excesses in AGN.

Our spectral fits did not combine both the UV and X-ray spectra. There
is about one order of magnitude difference (about 9$\sigma$) between
the UV fluxes of the thermal disk inferred from the soft X-ray
spectral fits and the OM measurements
(Section~\ref{sec:results}). This can be due to several
factors. First, we have only used the simple MCD model, while a more
realistic disk spectral model is probably needed to fit broadband data
\citep[about three decades in frequency
here;][]{lone1989,rofami1992,vafa2009}. Secondly, the starlight or hot
gas emission might be significant in the UV. This is supported by the
little variability of the UV emission between XMM1 and the {\it Swift}
observation in 2011 February. Finally, some UV emission might come
from the reprocessing of the X-ray emission in the outer disk or
surrounding gas.

\subsection{The Tidal Disruption Event Explanation}
\begin{figure}
\includegraphics{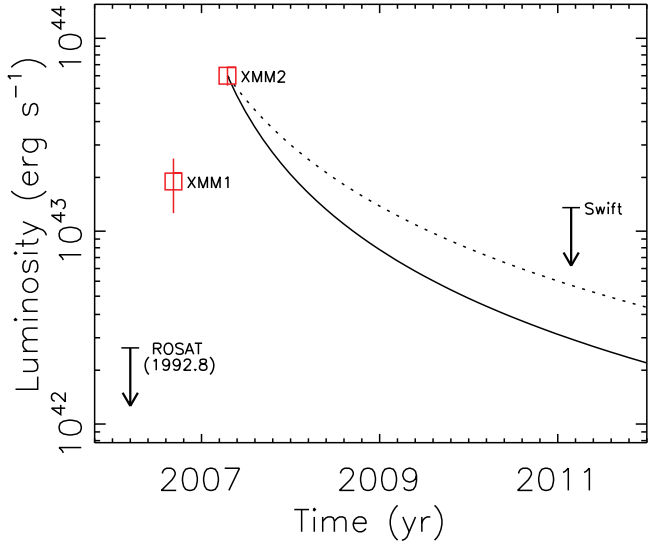}
\caption{The long-term luminosity curve inferred from X-ray spectral
fits. Arrows represent 3-$\sigma$ upper bounds. Note that the {\it
ROSAT} observation was made in 1992 October. The solid curve is ${\rm
Luminosity}=3.45\times 10^{43}[({\rm Time}-2006.60 {\rm yr})/(1 {\rm
yr})]^{-5/3}$ erg s$^{-1}$, and the dotted curve is ${\rm
Luminosity}=7.57\times 10^{43}[({\rm Time}-2006.18 {\rm yr})/(1 {\rm
yr})]^{-5/3}$ erg s$^{-1}$. \label{fig:longtermlc}}
\end{figure}
We show above that the X-ray emission in XMM1 and XMM2 can be best
explained as coming from a thermal disk around a SMBH with a mass of
$\sim$$10^5$--$10^6$ \msun. The transient nature of the source makes
it a great tidal disruption event candidate. This is further supported
by the extreme softness of its X-ray spectra and the inactivity of the
nucleus of the candidate host galaxy IC 4765-f01-1504. The inactivity
of IC 4765-f01-1504, i.e., not an AGN, is indicated by its lack of
significant optical emission lines (Figure~\ref{fig:optspec}). As our
new optical spectrum was made four years after the flare, this also
implies no detection of optical emission lines due to the flare at
this stage. The 2MASS IR colors (Table~\ref{tbl:counterpart}) also put
this galaxy in the region occupied by inactive galaxies in the
color-color diagram \citep[e.g.,][]{hyal1982}. The estimated BH mass
allows the tidal disruption of a solar-type star to be observable
\citep{lioz1979, re1988}. The luminosity reached 6.4$\times$10$^{43}$
erg~s$^{-1}$, which is about the average seen in other candidates
\citep{ko2002, essafr2007, gehece2009}. However, no previous
candidates had soft X-ray spectra with such high quality during the
peak of the flare to allow for the detailed spectral studies here.

One main feature of tidal disruption events is the temporal
evolution. Such events are predicted to rise on timescales of several
months and decay on timescales of months/years \citep{lioz1979,
re1988, re1990}. The decay approximately follows $L\propto (t-t_{\rm
D})^{-5/3}$, where $t_{\rm D}$ is the tidal disruption time, for most
candidates \citep{mauler2010,essako2008,hageko2004,koba1999},
consistent with the theory. XMM1 is fainter than XMM2, and they are
separated by 211 days. If our source is due to tidal disruption, XMM1
should probably be in the rising phase, and XMM2 in the decay
phase. We plot the luminosity curve in
Figure~\ref{fig:longtermlc}. Two decay curves following $L\propto
t^{-5/3}$ are also plotted, with XMM2 assumed to be in the decay. The
dotted and solid curves assume $t_{\rm D}$ to be at half a year and
one month before XMM1, respectively. These assumptions are reasonable,
as the minimum period for the material to return to the SMBH (so as to
accrete) after disruption is on the order of a month
\citep{re1988,evko1989}. These curves predict that the source
luminosity in the {\it Swift} observation in 2011 February should be a
factor of about 20 less than XMM2. Our detection limit is consistent
with this. In the above, we concentrate only on the decay, as the
connection from the rise to the decay is uncertain. The rise could
probably take months, and there could be a short period of several
months when the luminosity is maintained near the Eddington limit
\citep{re1988,re1990}. Future sensitive X-ray observations in the
decay are needed to constrain the long-term evolution.

We see that our source showed large fast X-ray variability, with an
rms of $\sim$21\% (Section~\ref{sec:pds}) in both \xmm\
observations. If they are really dominated by the thermal disk
emission, the above rms value is high, compared with the values of
$\lesssim$5\% typically seen in the thermal state of neutron-star (NS)
or BH low-mass X-ray binaries \citep{lireho2007,remc2006,mcre2006}. We
note that our rms was integrated over two decades in frequency, the
same as the above studies. Besides, in low-mass X-ray binaries, the
power of the thermal disk typically scales with the frequency as
$\nu^{-1}$ \citep{mcre2006}, while the power for our source is
steeper, scaling approximately as $\nu^{-1.7}$. Large fast variability
and steep power can be seen sometimes in the thermal state of X-ray
binaries, such as the flaring branch of bright NSs \citep{hovajo2002},
which was ascribed to some local instability in the inner disk by
\citet{lireho2009}. For the case of our source, we have shown in
Section~\ref{sec:pds} that its fast variability can be explained as
due to fast variations in the mass accretion rate, which, in the
context of a tidal disruption event, could reasonably be ascribed to
shocks during drastic compression and distortion of the stellar
material \citep{re1988}. This does not occur in low-mass X-ray
binaries, in which the mass is transferred through the Roche
lobe. Thus we speculate that the large fast variability and the steep
power in the thermal state is intrinsic to tidal disruption
events. Among the previous tidal disruption event candidates with
thermal X-ray peaks, the peak of \object{NGC 5905} was the best
observed. It showed an increase of a factor of $\sim$3 during the
peak, but it was over four days \citep{bakoda1996}, and the data
quality was not high enough to investigate the variability on much
shorter timescales.

\subsection{Comparison with AGN}

\object{2XMMi~J184725.1-631724} is unlikely to be an AGN from the lack
of bright optical emission lines. In the following we briefly compare
the X-ray properties of our source with ultrasoft AGN to show their
similarities and differences. NLS1 galaxies typically show the
steepest soft-X-ray spectra among AGN and have a typical photon index
of around 3 \citep{bobrfi1996,grkole2010}. The simple fits of the XMM1
and XMM2 spectra using the PL model indicate an extreme softness of
our source that is not seen in NLS1s.

We estimate the optical-to-X-ray spectral slope $\alpha_{\rm ox}$
using the flux densities in the rest frame of 2500 \AA~and 2 keV
\citep{taavbr1979}. We obtain $\alpha_{\rm ox}$=1.76$\pm$0.03. The
luminosity density in the rest frame of 2500 \AA~is
$l_{2500}$=(1.48$\pm$0.24)$\times$10$^{28}$ erg s$^{-1}$
Hz$^{-1}$. With this value of $l_{2500}$, the value of $\alpha_{\rm
ox}$ should be $\lesssim$1.3, based on the sample of 92 Seyfert 1
galaxies in \citet[][their Figure 15]{grkole2010}. In fact
\object{2XMMi~J184725.1-631724} has about the highest value of
$\alpha_{\rm ox}$ and the lowest value of $l_{2500}$ and is an
outlier, compared with their sample. We note, however, the possible
large uncertainty of $\alpha_{\rm ox}$ for our source, whose star
light contamination might be large, as discussed above.

There have been several ultrasoft AGN claimed in the literature. The
NLS1 galaxy WPVS 007 has the softest X-ray spectrum among AGN detected
during the {\it ROSAT} All-Sky Survey, with $\Gamma_{\rm PL}$$\sim$8
if fitted with a PL or $kT$$\sim$$20$ eV with a BB
\citep{grbema1995}. It can be explained as emission from the inner
disk \citep{grbema1995}, but the quality of the data and the lack of
simultaneous observations above 2.4 keV could make its soft excess due
to the presence of a warm absorber \citep{grleko2008}. The narrow-line
quasar PHL 1092 has \xmm\ observations, and the X-ray spectra are
steep ($\Gamma_{\rm PL}$$\sim$4--5), but the PL fit is not good when
it is bright \citep{gabobr2004, mifabr2009}. Its bright spectrum in
2003 was fitted with a model of MCD+PL plus an absorption line and a
reflection component by \citet{gabobr2004}; its soft excess was
ascribed to the MCD component ($kT$=114$\pm$4 eV), as in our
study. The NLS1 galaxy 1H 0707-495 also shows an intense soft
excess. Its X-ray spectra from \xmm\ show $\Gamma_{\rm PL}$$\sim$3.8
\citep{bofasu2002}. They were fitted with a model of BB+PL plus a
reflection component by \citet{fazoro2009}. From this model, both the
thermal disk emission and the reflection contribute to the soft
excess, but the latter dominates. We note that the above ultrasoft AGN
show large short-term variability factors of a few and/or long-term
variability factors of a few hundred, similar to our source. Thus the
variability of our source is not extreme compared with NLS1s, but the
softness of its X-ray spectra is hardly challenged. We note that the
soft component in our source cools in low states, which is not
generally seen in NLS1s \citep[e.g.,][]{mifabr2009}.

\subsection{Alternative Explanations}
We explore the possibility that \object{2XMMi~J184725.1-631724} is not
associated with the SMBH in IC 4765-f01-1504. If it is an
ultraluminous X-ray source (ULX) in this galaxy, it would have a
luminosity about one order of magnitude brighter than the brightest
ULX reported thus far, i.e., HLX-1 \citep{faweba2009}. However, as
derived above, both the spectral fits and the variability argument
imply that this source most probably has a mass of
$\sim$10$^5$--10$^6$ \msun, and it should be within 0.5 kpc (3-$\sigma$
error) of the galaxy center. Being so massive and so close to the
galaxy center, it seems unlikely to be a source other than the central
SMBH. Thus we deem that our source is not a ULX in IC 4765-f01-1504.

The other possibility is that \object{2XMMi~J184725.1-631724} is a
foreground Galactic source.  Following \citet{haru2009} and using the
0.2--10 keV absorbed flux of 1.92$\times$10$^{-12}$ erg s$^{-1}$
cm$^{-2}$ (Table~\ref{tbl:mcd+pl}), we have the X-ray to IR flux ratio
$f_{\rm X}/f_{\rm J}\sim 5$ for our source, making it unlikely to be a
coronally active star. These stars generally have $f_{\rm X}/f_{\rm
J}<0.03$ \citep{haru2009}.

The softness of the source with characteristic temperatures of a few
tens of eV makes it similar to the super-soft X-ray sources
\citep[SSS;][]{kava2006,gr2000}. This class of objects have BB
temperatures in the range 20--100 eV, which are about two orders of
magnitude lower than X-ray binaries containing an accreting NS or
BH. There are various types of SSS. One main class is cooling white
dwarfs \citep[WDs;][]{kava2006}.  \object{2XMMi~J184725.1-631724}
brightened by a factor of $\gtrsim$64, thus ruling out this
hypothesis. A large fraction of SSS can be interpreted as nuclear
burning of the hydrogen-rich matter on the surface of a white dwarf
that accretes matter from the companion, such as in the so-called
close binary super-soft sources and super-soft novae
\citep{kava2006,gr2000}. It seems unlikely that our source is such
based on the following considerations. In our Galaxy, there are only
about a dozen such sources detected since the {\it Einstein}
Observatory observations \citep{gr2000}, indicating a very low density
of such objects in the sky or a very low life duty cycle. The
probability for any of them lying in the direction of the center of a
galaxy is simply negligible. Besides, these objects are typically
observed at luminosities of $\sim$10$^{36}$--10$^{38}$ ergs~s$^{-1}$,
while our source has much lower luminosities, $<$10$^{34}$
ergs~s$^{-1}$, if it is 5 kpc away. These objects are mostly found at
distances of $<$5 kpc \citep{gr2000}.

\section{CONCLUSION}
\label{sec:conclusion}
\object{2XMMi~J184725.1-631724} is an ultrasoft X-ray transient source
with characteristic temperatures of a few tens of eV. It was bright in
two \xmm\ observations in 2006--2007, but was not detected in a {\it
ROSAT} pointed observation in 1992, implying a variation factor of
$\gtrsim$64 in the 0.2--10 keV absorbed flux. It was undetected again
in a {\it Swift} observation in 2011 February, implying a flux
decrease by a factor of $\gtrsim$12. It lies toward the center of the
galaxy IC 4765-f01-1504 at a redshift of 0.0353. No bright optical
emission lines were detected from this galaxy, making this source a
good tidal disruption event candidate. The fits to the two \xmm\
spectra using a thermal disk plus a weak hard component indicate that
the accretion disk luminosity appears to follow the $L\propto T^4$
relation and that the BH mass is around 10$^5$--10$^6$ \msun. The
source showed large fast variability in both \xmm\ observations, which
can be explained as due to fast variations in the mass accretion
rate. To further check whether this is a tidal disruption event,
future long-term X-ray monitoring is necessary to see whether it
follows the decay expected for a tidal disruption event.

\acknowledgments 

Acknowledgments: We thank the anonymous referee for the helpful
comments. We acknowledge the use of public data from the {\it ROSAT},
{\it Swift} and \xmm\ data archives, and the 2XMM Serendipitous Source
Catalog, constructed by the XMM-Newton Survey Science Center on
behalf of ESA. We want to thank the {\it Swift} PI Neil Gehrels for
approving our ToO request to observe the field of
\object{2XMMi~J184725.1-631724}. {\it Swift} is supported at PSU by
NASA contract NAS5-00136. The optical spectroscopy is based on
observations obtained at the Gemini Observatory which is operated by
the Association of Universities for Research in Astronomy, Inc., under
a cooperative agreement with the NSF on behalf of the Gemini
partnership: the National Science Foundation (United States), the
Science and Technology Facilities Council (United Kingdom), the
National Research Council (Canada), CONICYT (Chile), the Australian
Research Council (Australia), Minist\' erio da Ci\^ encia e Tecnologia
(Brazil) and Ministerio de Ciencia, Tecnolog\'{\i}a e Innovaci\' on
Productiva (Argentina). The observations were carried out as part of
program GS-2011A-Q-90. SAF acknowledges funding from the Australian
Research Council.

\end{document}